\documentclass[8pt,aps,prb,nofootinbib,twocolumn,latexsym]{revtex4}
\usepackage{amssymb}

\usepackage{latexsym}
\usepackage{amsmath}
\usepackage{graphicx}
\usepackage{subfigure}
\usepackage{color}
\usepackage{graphicx}

\begin{document}

\title{Suppression of Tunneling of Superconducting Vortices Caused by a Remote Gate: Example of an Extended Object Tunneling}
\author{K.~Michaeli and A.~M.~Finkel'stein}
\begin{abstract}
We discuss a recent experiment in which the resistance of a superconducting film has been measured in magnetic
field. A strong decrease of the superconducting film resistance has been observed when a metallic gate is placed
above the film. We study how the magnetic coupling between vortices in a thin superconducting film and electrons
in a remote unbiased gate suppresses the tunneling rate of the vortices. We examine two general approaches to
analyze tunneling in the presence of slow low-energy degrees of freedom: the functional-integral and scattering
formalisms. In the first one, the response of the electrons inside the metallic gate to a change in the vortex
position is described by the "tunneling with dissipation". We consider the Eddy current induced in the gate by
the magnetic flux of the vortex as a result of tunneling. In the second approach, the response is given in terms
of scattering of the electrons by the magnetic flux of the vortex in a way similar to the Aharonov-Bohm
scattering. A sudden change in the vortex position leads to the Orthogonality Catastrophe that opposes the
vortex tunneling. We show that the magnetic coupling between the vortices and the electrons inside the gate can
lead to a dramatic suppression of the vortex tunneling, restoring the superconducting property in accord with
the experiment.
\end{abstract}

\affiliation{Department of Condensed Matter Physics, The Weizmann Institute of Science, Rehovot 76100, Israel}
\maketitle

\section{Introduction}

The vortex motion in superconductors is a source of energy losses, destroying the perfect conductivity of the
superconductors. The dissipation is caused by non-superconducting electrons located inside the vortex
core.~\cite{Bardeen1965} The pinning potential created by impurities opposes the motion of vortices. This
potential results from the action of the impurities on the vortex core averaged over the area of the core.
Therefore, the pinning potential has minima typically separated by a distance of order of the coherence length
$\xi$.~\cite{Larkin1970,Larkin1972} Vortices may change their positions either by thermal
activation~\cite{AndersonKim} or by quantum tunneling between the potential minima at low enough
temperatures.~\cite{Glazman1992,Ephron1996,Mooij1996,Kogan2005} In this context, the observation of a strong
decrease of the resistance when an unbiased metallic gate is placed above an amorphous superconducting
film~\cite{Mason2002} is of great interest. We believe that this experiment provides a strong argument that at
low temperature the motion of vortices is indeed realized by quantum tunneling (in the experiment described in
Ref.~\onlinecite{Mason2002} it occurs at $T\lesssim0.1K$). If so, the tunneling of a vortex in a thin
superconducting film is a unique example of tunneling of a very extended "object".

In the discussed experiment, the resistance of a superconducting film has been measured at various magnetic
fields, both with and without a gate. In the absence of a gate, in magnetic fields lower than the critical one
($H<H_{C2}$) , the resistance initially decreases with lowering the temperature, but eventually saturates at a
finite value. The saturation indicates the possibility of vortex tunneling. When an unbiased metallic gate is
placed above the superconducting film the resistance reduces significantly with no indication of saturating at a
finite resistance when $T\rightarrow0$. Remarkably, the effect of the gate becomes noticeable at the same
temperatures where the resistance of the ungated film starts to saturate. It is worth mentioning that the gate
is separated from the film by an oxide layer of $160{\AA}$. Therefore, the film is thermally isolated from the
gate ruling out the possibility that the saturation of the resistance in the ungated film can be attributed to
heating. In this paper we identify the mechanism causing the suppression of the vortex motion in the presence of
the gate, which is effective in the tunneling regime only. The fact that the finite resistance at low
temperatures has been eliminated by placing a remote isolated gate strongly confirms the tunneling nature of the
vortex motion.

In the experiment of Ref.~\onlinecite{Mason2002} the film thickness is $a\approx30\AA$ and the gate thickness is
$d\approx400\AA$. An important feature indicating that the gate and the film are well separated, is that the
superconducting transition temperature, $T_{c}$, as well as the critical magnetic field, are practically
unchanged by adding the gate. Since the gate does not affect the superconductivity at $T\approx T_{c}$ when it
is the weakest, its influence on the superconducting properties, like the energy gap, at lower temperatures can
be ignored. One should also keep in mind that the gate does not influence the resistance of the film when the
vortex motion is thermally activated. In view of all of the above, we concentrate only on examining the
influence of the gate on the vortex tunneling rate. We assumed that the superconducting film and the gate are
magnetically coupled via the magnetic field of the vortices that pierces through the gate.

We employ here the following strategy. We accept the tunneling of vortices
at low temperatures as an established experimental fact. We do not try to
calculate the tunneling rate of the vortices. Instead, we concentrate on how
the response of the electrons inside the gate to a change of the vortex
position suppresses the tunneling rate. With this question in mind, in the
complex problem of the vortex tunneling, we wish to isolate the effect
induced by the gate.

\begin{figure}[pt]
\begin{flushright}\begin{minipage}{.5\textwidth}  \centering \subfigure[]{
        \label{fig:metal} %% label for first subfigure
        \includegraphics[width=0.45\textwidth]{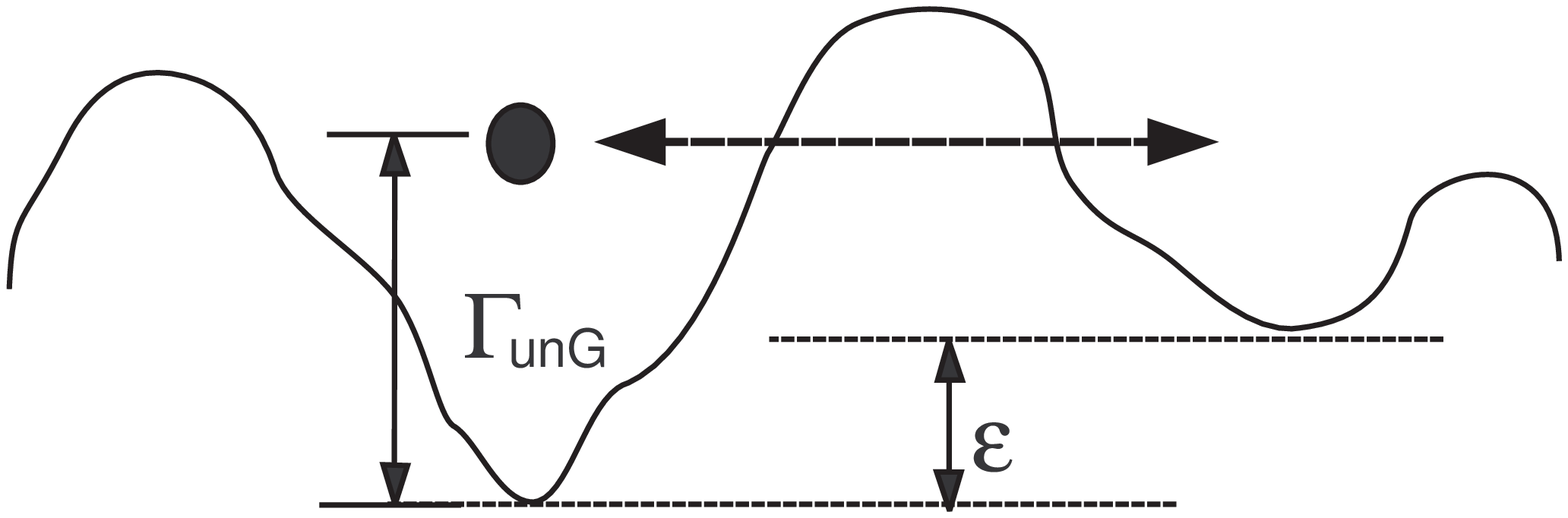}} \hspace{0.05in}%
        \subfigure[]{
        \label{fig:insulator} %% label for second subfigure
        \includegraphics[width=0.45\textwidth]{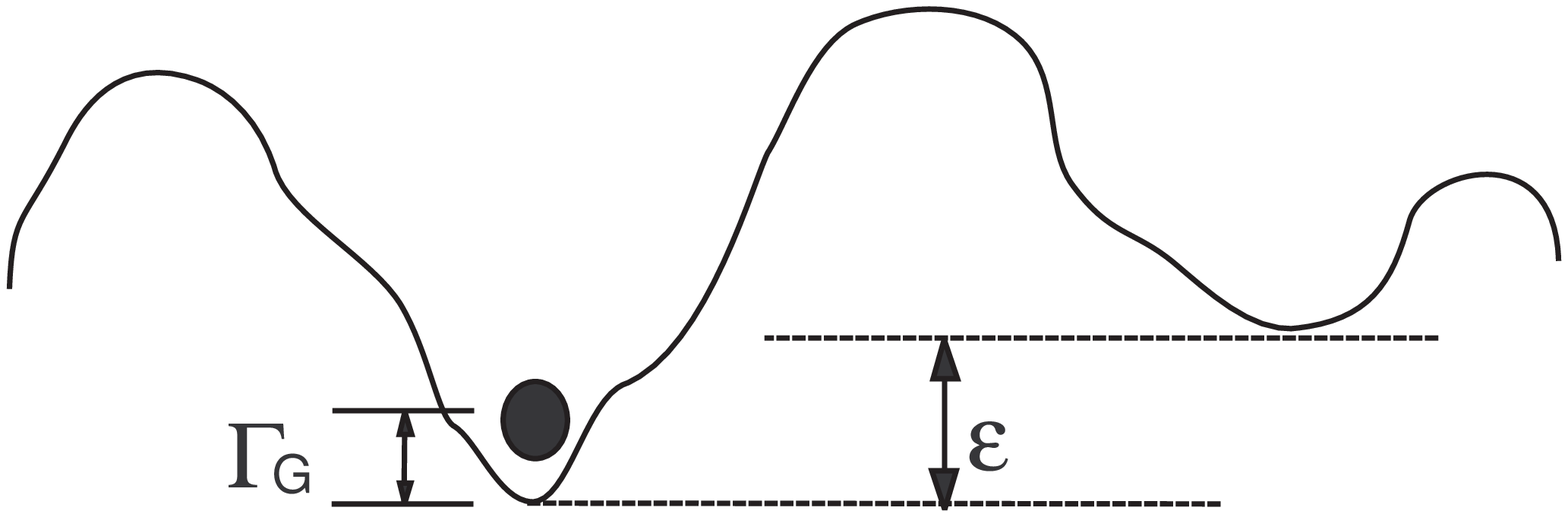}}
          \caption[0.4\textwidth]{\small The vortex in the effective potential
          landscape is represented by a hopping "particle".
        (a) The tunneling rate in the absence of the gate $\Gamma_{unG}$ exceeds the energy difference between the potential minima $\varepsilon$; the vortices are in a "metallic" phase
        (b) The tunneling rate is reduced by the gate to $\Gamma_{G}<\varepsilon$, and the system of vortices becomes an "insulator".} \label{fig:MI}
\end{minipage}\end{flushright}
\end{figure}

In fact, little is known about the motion of vortices at low temperatures.~\cite{Fisher1991} Fortunately, for
studying the role of the gate, it is sufficient to assume that the change in the vortex position is a discrete
tunneling event. This can be a tunneling of a single vortex, a bundle of vortices, or topological defects such
as dislocation pairs in the case of a vortex lattice (or a glass state). Phenomenologically, the change in the
vortex position can be described by the hopping Hamiltonian
\begin{equation}
H=\sum_{i}\varepsilon _{i}a_{i}^{+}a_{i}+\sum_{\left\langle i,j\right\rangle }\left( \Gamma
_{ij}a_{i}^{+}a_{j}+h.c.\right). \label{eq: Quantum Tunneling}
\end{equation}%
According to the standard criterion of the Metal-Insulator transition~\cite{Mott}, the ratio of the variation of
the potential minima $\varepsilon =\left\langle \varepsilon _{i}\right\rangle $ to the typical value of the
tunneling rates $\Gamma =\left\langle \Gamma _{ij}\right\rangle $ specifies whether the vortices are itinerant
or localized. The finite resistance at low temperature in the absence of the gate indicates that the vortices
are mobile, i.e., the tunneling rate in the ungated film $\Gamma_{unG}>\varepsilon $. The dissipationless nature
of the superconducting film is revived when the tunneling rate is reduced by the gate to
$\Gamma_{G}<\varepsilon$; see Fig.~\ref{fig:MI} which illustrates the two cases. Thus, we interpret the
experiment~\cite{Mason2002} as a transition from "metallic" to "insulating" phases in a system of tunneling
vortices that is induced by the gate. (Because of the strong interaction between the vortices, it may be enough
to have a fraction of vortices with a suppressed probability of tunneling acting as the pinning centers for the
entire ensemble of vortices.)

The fact that the tunneling of the vortices can be blocked by placing a gate above the film indicates that the
tunneling event gives rise to a dramatic response of the electrons inside the gate. This response can be
analyzed in terms of low-energy electron-hole excitations "decorating" the tunneling event. One may consider the
cloud of virtual excitations as part of the tunneling process that lasts long after the change in the vortex
positions occurs. In the present work we concentrate on the response of the electrons to the tunneling that can
lead to the strong suppression of the tunneling rate.

One may better understand the specifics of tunneling with the participation of low-energy degrees of freedom,
i.e., in the presence of dissipative environment, using the picture given by Iordanskii and one of the
authors.~\cite{iordanskii1973} Originally it described the quantum formation of a nucleation center in the decay
of a metastable macroscopic state, but it could also provide a general perspective. The quantum nucleation is an
example of a tunneling process in which a large number of degrees of freedom participate. Alternatively, this
kind of process can be treated as a tunneling of an artificial "particle" in a multi-dimensional space. When
low-energy degrees of freedom are involved in the process of quantum nucleation, the nucleation develops in two
stages.~\cite{iordanskii1973} Namely, the motion of the "particle" in this multi-dimensional space along the
trajectory minimizing the imaginary time action consists of fast and slow stages.

The slow stage appears because of the long time needed for the slow low-energy degrees of freedom to adjust
themselves to the new state of the fast degrees of freedom.~\cite{Iord F-comment} This time is much longer than
needed for fast the degrees of freedom to complete the tunneling. That is why the tunneling has to develop in
two stages. It has been shown in Ref.~\onlinecite{iordanskii1973} that despite the fact that the low-energy
degrees of freedom yield only a small contribution to the energy of the barrier, their participation in the
tunneling process increases parametrically the overall tunneling time. This results in a large increase of the
action and, correspondingly, in the strong suppression of the tunneling rate.~\cite{Comment}

The described picture of changing a quantum state in the presence of low-energy degrees of freedom is rather
typical for condensed matter systems. In the course of the fast stage of the process a quantum mechanical object
changes its state (position, spin projection, phase of the Josephson junction, etc). The accompanying slow
degrees of freedom act as an environment for the fast degrees of freedom. In the discussed problem of tunneling
in the gated superconducting film, the tunneling of the vortex from one potential minimum to another corresponds
to the fast stage. During the slow stage, the electrons inside the gate adjust their state to the new position
of the vortex.

The ensemble of electron-hole pairs in the gate represents the low-energy degrees of freedom of the environment.
The environment produces the most significant effect on tunneling at large time differences when the slow
degrees of freedom have enough time to develop. Naturally, the effect is the strongest when the tunneling occurs
between states that are almost degenerate. In the latter case the long time response can considerably reduce or
even block the tunneling~\cite{Leggett1987,Schmid1983} (this statement is often formulated in terms of the
dissipative Quantum Phase Transition~\cite{Zaikin2002}). With this in mind, we concentrate only on the slow
stage that develops when the tunneling of the vortex degrees of freedom is mostly accomplished without
specifying how the fast stage develops.

In this paper, we examine two different approaches to describe the response of the electrons inside the gate to
the change of the vortex position. In  Secs.~\ref{sec:Foucault_currents} and~\ref{sec:EOM} we consider the
dissipative Eddy (Foucault) currents induced in the gate which continue long after the vortex changed its
position. We formulated the effect of the gate on the vortex tunneling in terms of the effective action of a
vortex. We integrate out the environmental degrees of freedom inside the gate, and obtain the dissipative term
in the effective action. This allows us to consider the vortex tunneling in the gated superconducting film in
the context of the well-known problem of tunneling in the presence of a dissipative
environment~\cite{Caldeira1981}, Sec.~\ref{FCDiscussion}. Alternatively, one can analyze the slow stage in terms
of the scattering of electrons. In Sec.~\ref{sec:OC} we describe the elastic scattering of the gate electrons on
the vector potential of the magnetic field of the vortex. The zero overlap between the states of the electrons
before and after the change of the scattering potential is known generally as the Orthogonality
Catastrophe~\cite{AndersonOC1967}(OC). We show that the OC caused by the change in the vortex position,
effectively suppresses the vortex tunneling. The novel element here~\cite{KM2006} is that the \textit{OC} is
connected to the Aharonov-Bohm effect.~\cite{Aharonov1959,Aharonov1984} We obtained that the \textit{OC} is
significantly more effective in suppressing the tunneling rate than the Eddy current. In
Sec.~\ref{sec:discussion} we discuss the relation between the two approaches and the peculiarities of the
tunneling of an extended object, such as a vortex in a thin superconducting film.

\section{Eddy Currents Inside the Gate}

\label{sec:Foucault_currents}

The magnetic field of the vortex inside a superconducting film is similar to the magnetic field of a solenoid
with a radius equal to the magnetic penetration depth $\lambda $ which can be very large.~\cite{Pearl1964}
Outside of the film, the magnetic field decays as a function of the height and deflects into the radial
direction~\cite{Abrikosov,Carneiro2000}:
\begin{equation}
\mathbf{A}_{vor}(\mathbf{r,}z\mathbf{;R})=\frac{\alpha \Phi _{0}}{\lambda }\int \frac{d^{2}q}{(2\pi
)^{2}}e^{i\mathbf{qr}-q|z|}\frac{i\mathbf{q}\times \hat{z}}{q^{2}(\lambda ^{-1}+2q)}.  \label{eq:vortex}
\end{equation}
Here $\mathbf{r}$ is the radial vector in cylindrical coordinates with the origin at the vortex center
$\mathbf{R}$. Since a vortex in a film can move only in the $x-y$ plane, $\mathbf{R}$ is a two dimensional
vector. The parameter $\alpha =1/2$ is the total flux of a vortex in a superconductor measured in units $\Phi
_{0}=2\pi \hbar c/e$.

In the case of a thin superconducting film, a Pearl vortex is a macroscopically large object. The penetration
depth $\lambda $ is~\cite{Pearl1964}:
\begin{equation}
\lambda =\lambda _{3D}^{2}/a\;\qquad \lambda \gg a,
\label{eq:penetration depth}
\end{equation}%
where $a$ is the thicknesses of the superconducting film, $\lambda _{3D}$ is the penetration depth in a
disordered bulk superconductor. In Eq.~(\ref{eq:penetration depth}), $\lambda _{3D}\sim \lambda _{L}(\xi
_{0}/\ell )^{1/2}$, where $\lambda _{L}=(mc^{2}/4\pi ne^{2})^{1/2}$ and $\xi _{0}$ are the London length and the
coherence length of a clean superconductor; $\ell $ is the mean free path.

\begin{figure}[]
\begin{flushright}\begin{minipage}{0.5\textwidth} \centerline{
    \includegraphics[width=0.9\textwidth]{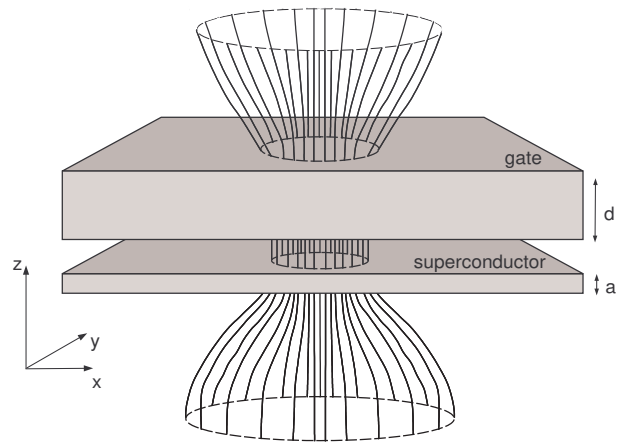}}
\caption{\small Superconducting film magnetically
    coupled to a metallic gate. The magnetic field of the
vortex pierces through the gate. In the experiment of Ref.~\onlinecite{Mason2002} $d\approx400{\AA}$,
$a\approx30{\AA}$, and the insulating layer between the gate and the film is $160{\AA}$ thick.}
\label{fig:vortex_field}
\end{minipage}\end{flushright}
\end{figure}

In the experiment of Ref.~\onlinecite{Mason2002} the gate thickness $d$ is much smaller than $\lambda$. Because
of the exponential decay of the Fourier components in the $z$-direction, see Eq.~(\ref{eq:vortex}), momenta that
contribute mostly are limited to $q\lesssim1/d$. More accurately, $q\lesssim \min \{d^{-1},\delta r^{-1}\}$,
where $\delta{r}$ is the typical distance that a vortex has to tunnel, $\delta{r}\sim\xi$. To avoid unnecessary
complications, we ignore the space between the superconducting film and the gate since it is considerably
smaller than the gate thickness, see Fig.~\ref{fig:vortex_field}. Then, for small momenta that we are interested
in, the deflection of the magnetic field in the space between the film and the gate can be neglected.

To describe the tunneling of a vortex in the presence of a gate one has to
deal with an imaginary time action:
\begin{equation}
S=S_{sc}+S_{gate}+S_{int}.  \label{eq:S}
\end{equation}
In what follows we discuss each term in the action $S$.

The term $S_{sc}$ is an action of the superconducting film in the absence of
the gate. Since we are not trying to solve the problem of the vortex
tunneling in full scale, but are interested only in the effect of the gate
on the tunneling rate, this part of the action is not specified.

In writing $S_{gate}$ which describes the dynamics of the electrons in the gate one should keep in mind the
following argument. The charge and current densities relevant for the long tail response of electrons inside the
gate to the tunneling of the vortex are characterized by large length and time scales.\emph{\ }Therefore, their
dynamics can be described macroscopically. Since the deviations of the charge and current densities from their
equilibrium values are small, the action that describes the dynamics of their fluctuations should be consistent
with the Fluctuation-Dissipation Theorem (FDT). (Examples of such an approach can be found in the calculation of
the dephasing time of the cooperons due to the electromagnetic fluctuations~\cite{Altshuler1982}, and also in a
macroscopic calculation of the zero bias anomaly.~\cite{Levitov1997})

The current in the gate has two contributions. One is the Ohmic response to the electric field, while the other
one is the diffusive current from the gradients of the density; $\mathbf{J}(\mathbf{r},\tau
)=\mathbf{J}_{ohmic}(\mathbf{r},\tau )-D\boldsymbol{\nabla}\rho (\mathbf{r},\tau )$. The fluctuations of the
charge and current densities can be expressed through the correlation function of the Ohmic part of the current
written in terms of the Matsubara frequency as follows:
\begin{align}
\hat{K}_{i,j}^{-1}&(\mathbf{k},i\omega _{n})=\langle \mathbf{J}_{ohmic}^{i}(%
\mathbf{k},i\omega _{n})\mathbf{J}_{ohmic}^{j}(-\mathbf{k},-i\omega
_{n})\rangle  \label{eq:FDT} \\
& =\sigma _{i,j}(\mathbf{k},\omega _{n})|\omega _{n}|+\sigma _{i,i^{\prime }}(\mathbf{k},\omega
_{n})D_{j,j^{\prime }}(\mathbf{k},\omega _{n})k_{i^{\prime }}k_{j^{\prime }}. \notag
\end{align}
Here the diffusion constants tensor $\hat{D}$ and the conductivity $\hat{\sigma}$ are connected through the
Einstein's relation $\hat{\sigma}=e^{2}(dn/d\mu )\hat{D}$. The gate, being a simple homogeneous metal, is
adequately described by the Drude formula. We assume that the external magnetic field is classically weak,
$\omega _{c}\tau \ll 1,$ so that we can ignore the Hall conductivity in the gate; $\tau $ is the mean free time
in the gate. Since we are interested in the low frequency and long wave length behavior only, we take the
conductivity $\sigma $ to be constant.

Following the above arguments, one can write the gate action $S_{gate}$ for
the charge and current densities inside the gate in the Matsubara time:
\begin{align}
S_{gate}& =\frac{1}{2}\int_{0}^{\beta }d\tau _{1}d\tau _{2}\int
{d}\mathbf{r_{1}}d\mathbf{r_{2}}L_{FDT}[\mathbf{r_{1}},\tau _{1};\mathbf{r_{2}},\tau
_{2}]  \label{eq:Gate_Action} \\
& +\int_{0}^{\beta }d\tau \int {d}\mathbf{r}[\mathcal{L}_{cont}+\mathcal{L}%
_{Max}]  \notag
\end{align}
The first term in the above action describes the charge and current density
fluctuations in accordance with the FDT:
\begin{align}
L_{FDT}& =\left[ \mathbf{J}(\mathbf{r_{1}},\tau _{1})+\hat{D}\boldsymbol{\nabla}\rho (\mathbf{r_{1}},\tau
_{1})\right] \hat{K}(\mathbf{r_{1}}-\mathbf{r_{2}},\tau _{1}-\tau _{2}) \\
& \left[ \mathbf{J}(\mathbf{r_{2}},\tau _{2})+\hat{D}\boldsymbol{\nabla}\rho (\mathbf{r_{2}},\tau _{2})\right].
\notag
\end{align}
Since the fluctuations of the current and density are not independent, the
second term in Eq.~(\ref{eq:Gate_Action}) imposes the charge continuity
condition with the use of the Lagrange multiplier $\phi $:
\begin{equation}
\mathcal{L}_{cont}=\phi (\mathbf{r},\tau )\left( i\dot{\rho}(\mathbf{r},\tau
)+\boldsymbol{\nabla}\mathbf{J}(\mathbf{r},\tau )\right)
\end{equation}%
The term $\mathcal{L}_{Maxwell}$ describes the interactions of the electromagnetic fields and the charge and
current densities in the gate in a way that reproduces the Maxwell equations:
\begin{align}
\mathcal{L}_{Max}& =i\frac{1}{c}\left[ \mathbf{J}\mathbf{A}_{ind}+\rho
\varphi _{ind}\right] \\
& +\frac{1}{8\pi }\left[ \left( -\boldsymbol{\nabla}\varphi _{ind}-\frac{i}{c%
}\mathbf{\dot{A}}_{ind}\right) ^{2}+(\boldsymbol{\nabla}\times \mathbf{A}%
_{ind})^{2}\right].  \notag
\end{align}
The factor $i$ in the coupling of the electromagnetic fields with the charge and current densities appears
because of the imaginary time.~\cite{Popov} The plus in the magnetic field term (which is a consequence of the
$i$-factor) is needed to get the repulsive sign in the Ampere interaction of currents, like in the Coulomb
interaction of charges. As usually, the total potentials are the sum of the external and induced potentials; in
the discussed system the external source is the field of the vortices.

The last term in the action~(\ref{eq:S}), $S_{int}$, describes the
connection between the superconducting film and the gate. The current and
charge densities in the gate interact with the vector and scalar potentials
created by the superconducting film:
\begin{eqnarray}\label{eq:S_int}
 S_{int}&=&i\int_{0}^{\beta }d\tau \int {d}\mathbf{r}\left\{\frac{1}{c}\mathbf{J}(\mathbf{r},\tau )\cdot
\mathbf{A}_{sc}(\mathbf{r},\tau ) \right.  \nonumber \\
 && \hspace{32mm}+\left. \rho (\mathbf{r} ,\tau ){\varphi }_{sc}(\mathbf{r},\tau ) \phantom{\frac{1}{c}\mathbf{J}(\mathbf{r},\tau )\cdot
\mathbf{A}_{sc}(\mathbf{r},\tau
)}\!\!\!\!\!\!\!\!\!\!\!\!\!\!\!\!\!\!\!\!\!\!\!\!\!\!\!\!\!\!\!\!\!\!\!\!\!\!\!\!\!\!\!\!\!\!\!\!\!\!\right\}
\end{eqnarray}

We are interested in the limited problem of the long tail response of electrons inside the gate that develops
when the tunneling of the vortex degrees of freedom is mostly accomplished. Therefore, we consider large enough
$\tau $ when one can assume that the deformation of the field $\mathbf{A}_{sc}(\mathbf{r},\tau )$ appearing
during the process of vortex tunneling has been already relaxed. [By deformation we mean the deviation of
$\mathbf{A }_{sc}(\mathbf{r},\tau )$ from the field of the ''rigid'' vortex centered at $\mathbf{R}(\tau )$, as
it is given by Eq.~(\ref{eq:vortex}). Here $\mathbf{R}(\tau )$ denotes the position of the vortex at time $\tau
$.] With this in mind, we put in $S_{int}$
\begin{equation}
\mathbf{A}_{sc}(\mathbf{r},\tau )=\mathbf{A}_{vor}(\mathbf{r};\mathbf{R}(\tau )).  \label{eq:A=A}
\end{equation}%
Furthermore, we ignore the scalar potential, $\varphi _{sc}(\mathbf{r},\tau )=\varphi
_{vor}(\mathbf{r};\mathbf{R}(\tau ))=0$, relying on the known fact that the redistribution of the charge density
around the vortex core is negligible. We still have to justify our treatment of the electromagnetic field
$\mathbf{A}_{sc}$ as a given external field in the analysis of the long tail response of the gate electrons. We
will see in the next section that the magnetic field created by the low-frequency components of the dissipative
Eddy currents is much smaller than the field of the vortex, and can perturb the superconducting film only
weakly. Therefore, the gate does not provide a substantial feedback effect to the superconducting film during
the slow stage.

To get the response of the environment on the vortex motion, one has to integrate out the gate degrees of
freedom. Since $S_{gate}$ and $S_{int}$ are quadratic in the charge and current densities, this immediately
results in
\begin{equation}
S_{env}=\frac{i}{2}\int_{0}^{\beta }d\tau \int dr\frac{1}{c}\mathbf{J}_{cl}(\mathbf{r},\tau )\cdot
\mathbf{A}_{vor}(\mathbf{r},\mathbf{R}(\tau )). \label{eq:S_env}
\end{equation}%
The current $\mathbf{J}_{cl}(\mathbf{r},\tau )$ has to be found by solving the classical equations of motion.
Since the current $\mathbf{J}(\mathbf{r},\tau )$ describes the long-time response of the electrons in the gate,
the term $S_{env}$ is non-local in time.

\section{The Solution of the Equation of Motion}

\label{sec:EOM}

As we stated above, the main contribution to the induced current emerges from the components of the field with
$q\lesssim1/d$ for which the depletion of the magnetic field of the vortex from the $z$ direction is negligible.
Therefore, we can ignore the current component in the $z$-direction and consider only the components that are
parallel to the plane. With this in mind, we decompose the current parallel to the plane of the gate into two
components: a longitudinal component along the two dimensional vector $q$ and a transverse component
perpendicular to it, $\mathbf{J}(\mathbf{q},\tau )=J^{\parallel }(\mathbf{q},\tau )\hat{\mathbf{q}}+J^{\perp
}(\mathbf{q},\tau )\hat{z}\times \hat{\mathbf{q}}$. In the following we use the Fourier transformation for the
in-plane coordinates only and keep the vertical coordinate separately. Then, the kernel in Eq.~(\ref{eq:FDT}) is
diagonal in the chosen basis. As it follows from Eq.~(\ref{eq:vortex}), the vector potential
$\mathbf{A}_{vor}(\mathbf{r})$ contains the transverse component only. Therefore, since the use of the FDT for
describing the slow stage implies a linear response to the external field, the longitudinal component
$J^{\parallel }$ cannot be generated. This is a direct consequence of the magnetic coupling between the
superconducting film and the gate. It is very different from the zero bias anomaly in which only a longitudinal
current is generated.~\cite{Levitov1997}

The variation of $S_{gate}+S_{int}$ with respect to the current and charge densities determines the equations of
motion. We use these equations for finding the transverse current:
\begin{align}
& \int {dz^{\prime }}K^{\perp }(\mathbf{q};z,z^{\prime };i\omega
_{n})J^{\perp }(\mathbf{q},z^{\prime },i\omega _{n})  \label{eq:J_Transverse}
\\
& =-\frac{1}{c^{2}}\int {dz^{\prime }}U(\mathbf{q},z-z^{\prime })J^{\perp }(\mathbf{q},z^{\prime },i\omega
_{n})-i\frac{1}{c}A_{vor}^{\perp }(\mathbf{q},i\omega _{n}).  \notag
\end{align}
The L.H.S of this equation is the total vector potential in the transverse direction calculated via its Ohmic
response: $J^{\perp }=(\sigma \omega _{n}/c)A_{total}^{\perp }$. The first term in the R.H.S is the vector
potential of the induced field and the second one is the external potential of the vortex. Neglecting
relativistic effects, the instant kernel $U$ can be written as the Fourier transform of the instant in time
Biot-Savart kernel $U(\mathbf{r}-\mathbf{r}^{\prime })=1/|\mathbf{r}-\mathbf{r}^{\prime }|$ with respect to the
in-plane momenta, $U(\mathbf{q},z-z^{\prime })=2\pi e^{-q|z-z^{\prime }|}/q$.

The above equation for the transverse current can be rewritten in the form:
\begin{equation}
\int {dz}^{\prime }L^{\perp }(\mathbf{q};z,z^{\prime };i\omega _{n})J^{\perp }(\mathbf{q},i\omega
_{n})=-i\frac{\sigma |\omega _{n}|}{c}A_{vor}^{\perp }(\mathbf{q},i\omega _{n}),  \label{eq:J_TransverseL}
\end{equation}
where the kernel $L^{\perp }$ is
\begin{align}
& L^{\perp }(\mathbf{q};z,z^{\prime };i\omega _{n})= \\
& \left\{
\begin{array}{ll}
\delta (z-z^{\prime })+\frac{2\pi \sigma |\omega _{n}|}{c^{2}}{\
e^{-q|z-z^{\prime }|}}/q & \mbox{$0\leq{z,z'}\leq{d}$}; \\
\hspace{20mm}0 & \mbox{\hspace{5mm}\emph{otherwise}}.%
\end{array}%
\right.  \notag  \label{eq:Kernel_Ltran}
\end{align}
To find the transverse current one has to invert this kernel.

In analogy to the skin-effect, one can define a screening length $\delta(\mathbf{q},\omega
_{n})=1/\sqrt{q^{2}+4\pi \sigma |\omega _{n}|/c^{2}}$. [Usually, the skin-effect is discussed in the case of an
electromagnetic wave propagating normally to a surface of a metallic slab. The geometry of the problem studied
here is different as the magnetic field is normal to the slab surface, while the propagation is parallel to
it.\emph{\ }Still, the surface current appearing in the gate screens out the high-frequency components of the
field in the bulk of the slab.] The kernel $L^{\perp }$ can be inverted in the two limits: (i) for the
components of the electromagnetic field that are transparent for the gate, $d/\delta (\mathbf{q},\omega _{n})\ll
1;$ or (ii) for the components with $d/\delta (\mathbf{q},\omega _{n})\gg 1$ that are well screened by the
surface currents. These are the thin and thick gate limits, respectively.

In the thin gate limit, the change in the current along the $z$ direction is minor, and a reasonable
approximation (up to liner terms in $qd$) is to consider the current to be homogenous in the $z$ direction,
$I_{2D}=dJ^{\perp }(\mathbf{q},z,i\omega _{n})$. Then,
\begin{equation}
I_{2D}=-i\frac{\sigma ^{2D}}{1+\frac{2\pi \sigma ^{2D}|\omega _{n}|}{qc^{2}}}%
\frac{|\omega _{n}|}{c}A_{vor}^{\perp }(\mathbf{q},i\omega _{n})+O(qd).
\label{eq:J_tran_thin}
\end{equation}
One can observe that the expression for the current $I_{2D}$ is identical to the current in a two-dimensional
system with $\sigma ^{2D}=\sigma d$, and where $2\pi \sigma ^{2D}|\omega _{n}|/qc^{2}$ is the current screening
operator. [The term ''current screening operator'' is used here in analogy with the polarization operator. It
describes the screening of the transverse component of the time dependent vector potential by the induced
currents.]

At low frequencies such that $2\pi \sigma ^{2D}|\omega _{n}|/c^{2}<q\lesssim{d}^{-1}$ (this automatically
implies the thin gate limit), the current in the gate screens weakly the field produced by the vortex. The total
field is approximately just the field of the vortex, and the current is merely the Ohmic response to it:
\begin{equation}
I_{2D}=-i\frac{\sigma ^{2D}|\omega _{n}|}{c}A_{vor}^{\perp }(\mathbf{q},i\omega _{n}). \label{eq:Ohmic_thin}
\end{equation}
For higher frequencies, $2\pi \sigma ^{2D}|\omega _{n}|/c^{2}>q$, but still in the thin gate limit, the
situation is rather different. Since the effect of dissipation is stronger for better conducting gates, we are
interested in the case when $2\pi \sigma ^{2D}/c\gg 1.$ Then, there is a window $qc^{2}/2\pi \sigma
^{2D}<|\omega _{n}|<qc$ in which the current $I_{2D}$ obeys a London-like equation~\cite{Abrikosov},
$I_{2D}=-i\frac{qc}{2\pi } A_{vor}^{\perp }(\mathbf{q},i\omega _{n})$. Since the electrons instantly respond to
the potential, the contribution to the action in this limit is not of a dissipative character. Rather, it
provides a local in time term acting as an additional potential that should be added to $S_{sc}$. This kind of
contribution is not considered here.

For a thick gate, $\delta (\mathbf{q},\omega _{n})\ll d$, the limit of low frequencies does not exist for
$q\lesssim d^{-1}$. In this limit the screening length is equal to $\delta (\omega _{n})=c/\sqrt{4\pi \sigma
|\omega _{n}|}$. At such high frequencies the current flows in the reduced volume (which effectively is a thin
slab of width $\delta (\omega _{n})$) as follows:
\begin{eqnarray}
&&J^{\perp }(\mathbf{q},z,i\omega _{n})=-i\frac{cq}{2\pi }\delta ^{-1}(\mathbf{q},\omega
_{n})e^{-z/\delta (\mathbf{q},\omega _{n})+qz}  \notag \\
&&\qquad \qquad \qquad \times A_{vor}^{\perp }(\mathbf{q},z,i\omega _{n})+O\left( e^{-d/\delta }\right).
\label{eq:J_tran_thick}
\end{eqnarray}%
Notice that the Fourier components $A_{vor}^{\perp }(\mathbf{q},z)$ decay exponentially on $z$ as $e^{-qz}$.
Therefore, the factor $e^{qz}$ in the solution above is canceled out leaving the induced surface current with
the $z$-dependence $e^{-z/\delta (q,\omega _{n})}$. Integrating Eq.~(\ref{eq:J_tran_thick}) in the $z$-direction
yields the London-like current $I_{2D}=-i\frac{qc}{2\pi }A_{vor}^{\perp }(\mathbf{q},i\omega _{n})$ exactly as
in the thin gate limit.

So far, we have ignored relativistic effects that appear at high frequencies when $|\omega _{n}|/c>{q}$. In this
case, one has to substitute $q$ by the relativistic combination $\sqrt{q^{2}+(\omega _{n}/c)^{2}}$. Using the
relativistic equations of motion, one can show that the current $\mathbf{J}_{cl}$ is like in the Ohmic regime
but with $\sigma ^{2D}$ replaced by $\sigma ^{2D}/(1+2\pi \sigma ^{2D}/c)$. In the limit $2\pi \sigma ^{2D}/c\gg
1$, the dissipation caused by the Cherenkov's radiation~\cite{Fal'ko1989} corresponds to the effective
conductivity equal to $c/2\pi $. Still, the effect of the relativistic region is negligible in comparison to the
low-frequency contribution to $S_{env}$.

To conclude, let us come back to our assumption about the absence of a feedback from the gate to the
superconducting film. As we have showed above, the most significant contribution to $S_{env}$ originates from
the region ${2\pi |\omega _{n}|\sigma }^{2D}/({qc^{2}}){\ll }1$ when the Eddy current is in the Ohmic regime. In
this limit, the low-frequency components of the vortex field are poorly screened by the induced current in the
gate. Hence, the feedback from the gate to the superconducting film can be neglected.

We are ready to obtain the dissipative term in the action describing the effective response of the environmental
degrees of freedom on the vortex tunneling. Inserting the current $J^{\perp }$ from Eq.~(\ref{eq:Ohmic_thin})
into Eq.~(\ref{eq:S_env}), one gets

\begin{widetext}
\begin{align}\label{eq:Action-momentum}
S_{env}& =\frac{\sigma }{2\beta {c^{2}}}\int {d\tau _{1}}{d\tau _{2}} \sum_{n}|\omega _{n}|e^{-i\omega _{n}(\tau
_{1}-\tau _{2})}\int \frac{d^{2}q}{(2\pi )^{2}}dze^{i\mathbf{q}(\mathbf{R}(\tau _{1})-\mathbf{R}(\tau _{2}))}
\frac{-i\mathbf{q}\times\mathbf{A}_{vor}(-\mathbf{q},z)}{q}\frac{i\mathbf{q}\times\mathbf{A}_{vor}(\mathbf{q},z)}{q}.
\end{align}
Performing the sum over the frequencies and rewriting the action in terms of the magnetic field we get
eventually:
\begin{equation}
S_{env}=-\frac{\pi \sigma }{2\beta ^{2}{c^{2}}}\int d\tau _{1}d\tau _{2}\int
\frac{d^{2}q}{(2\pi )^{2}}dz\frac{e^{i\mathbf{q}(\mathbf{R}(\tau _{1})-%
\mathbf{R}(\tau _{2}))}}{\sin ^{2}\left( \frac{\pi }{\beta }(\tau _{1}-\tau
_{2})\right) }\frac{B_{vor}^{z}(-\mathbf{q},z)B_{vor}^{z}(\mathbf{q},z)}{%
q^{2}}\hspace{5mm}.  \label{eq:Action-time}
\end{equation}
\end{widetext}

\section{Tunneling in the presence of $S_{env}$}

\label{FCDiscussion}

For the purpose of illustration, let us compare $S_{env}$ with the action of
Refs.~\cite{Caldeira1981,Caldeira1983} for a particle moving in a dissipative environment:
\begin{equation}
S_{CL}=1/4\pi \int {d}\tau _{1}d\tau _{2}\eta \frac{(\mathbf{R}(\tau _{1})-%
\mathbf{R}(\tau _{2}))^{2}}{(\tau _{1}-\tau _{2})^{2}}.
\label{eq:Cal-Leg action}
\end{equation}%
This term results from integrating out the slow degrees of freedom of the environment. The long-time response of
these modes reveals itself through a non-local in time term in the action. If $\mathbf{R}(\tau )$ describes a
particle moving with a constant velocity, one can substitute $[\mathbf{R}(\tau )-\mathbf{R}(\tau ^{\prime
})]/(\tau -\tau ^{\prime })$ with the velocity $\dot{\mathbf{R}}$. Then the action $S_{CL}$ reduces to $1/2\pi
\int {d\tau _{1}d\tau _{2}}\eta \dot{\mathbf{R}}^{2}/2$ where the integrand is a reminiscent of the Rayleigh's
dissipation function. The Rayleigh function is used in the Euler-Lagrange equations to include
dissipation~\cite{Goldstein}:
\begin{equation}
\frac{d}{dt}\left( \frac{\partial \mathcal{L}}{\partial \dot{R}}\right) - \frac{\partial \mathcal{L}}{\partial
{R}}+\frac{\partial \mathcal{F}}{\partial \dot{R}}=0,
\end{equation}
where $\mathcal{F}=\eta \dot{R}^{2}/2$ is the Rayleigh function with a friction coefficient $\eta $. Since the
Rayleigh function enters the equation of motion without a time derivative (unlike the Lagrangian), the inclusion
of the dissipation into the action costs an additional time integration. Therefore, the corresponding term must
be non-local in time.

Under the same approximation of constant velocity, the action $S_{env}$ can be written in the coordinate
representation as:
\begin{equation}
S_{env}=\frac{1}{4\pi }\int {d^{2}r}dz\int {d}\tau _{1}d\tau _{2}\frac{\sigma }{c^{2}}\left[
\mathbf{v}_{vor}\times \mathbf{B}_{vor}(\mathbf{r,}z) \right] ^{2}.  \label{eq:Action-coordinates}
\end{equation}
The combination $\left[ \mathbf{v}_{vor}\times \mathbf{B}_{vor}(\mathbf{r})\right]/c $ is the electric field
created by a moving vortex. Then, the integrand in the action $S_{env}$ is merely the energy dissipation rate in
the gate caused by the vortex motion, $S_{env}=1/2\pi \int d^{2}rdz\int d\tau _{1}d\tau _{2}\sigma
{E}_{vor}^{2}(\mathbf{r},z)/2$. This expression is in full correspondence with the one obtained for a constant
motion of a particle in the presence of friction $\eta $.

We return now to Eq.~(\ref{eq:Action-time}) and analyze it for the case of tunneling between two minima
separated by a distance $\delta r$. We start with the integration over the coordinate $z$ and momentum
$\mathbf{q}$ using for the magnetic field $\mathbf{B}_{vor}$ the solution given by Eq.~(\ref{eq:vortex}). We get
in result
\begin{align}
S_{env}=\frac{\alpha ^{2}\sigma d}{16e^{2}\lambda ^{2}}& \int {d\tau _{1}}{d\tau _{2}}\frac{(\mathbf{R}(\tau
_{1})-\mathbf{R}(\tau _{2}))^{2}}{(\tau_{1}-\tau _{2})^{2}}  \label{eq:ActionEnv} \\
& \times \ln \left( \frac{d}{\lambda }+\frac{8\pi ^{2}\sigma ^{2D}{d}}{|\tau
_{1}-\tau _{2}|c^{2}}\right) ^{-1}.  \notag
\end{align}
The appearance of the log-factor is very natural if one recalls that we integrate the square of the magnetic
field and it is well known that $B_{vor}\sim 1/r$ at a distance $r$ from the center of the vortex, when
$r<\lambda $. The time dependence of the logarithm results from the fact that the integration over the momenta
is limited to the Ohmic regime, $|\omega _{n}|\ll c^{2}q/(2\pi \sigma ^{2D})$. The time dependence of the
logarithm is important because in thin superconducting films the ratio $d/\lambda $ can be very small. At large
time differences, which are essential for low temperatures, the logarithmic factor in the action becomes
$\ln\lambda /d$.

Following the standard Renormalization Group (RG) procedure one gets that the modified tunneling rate
$\Gamma_{G}$ is:
\begin{equation}
\Gamma _{G}(T)=\Gamma _{unG}\left( T\tau _{tun}\right) ^{K_{E}},
\label{eq:tunneling_frequency}
\end{equation}
where the exponent $K_{E}$ is equal to the dimensionless dissipation
coefficient
\begin{equation}
K_{E}=\sigma {d}\frac{(\alpha \delta {r})^{2}}{8e^{2}\lambda ^{2}}\ln \left( \frac{\lambda }{d}\right).
\label{eq:K{E}}
\end{equation}
In Eq.~(\ref{eq:tunneling_frequency}), $\tau _{tun}$ is the time of the under-barrier motion of the vortex in
the process of tunneling in the absence of the gate (i.e., the duration of the fast stage of the tunneling
discussed in the Introduction). The parameter $\tau _{tun}^{-1}$ acts as the high-energy cutoff because only
slow excitations that cannot follow adiabatically the tunneling particle contribute to the slow stage of the
tunneling process (it is assumed that $\tau _{tun}^{-1}>>T$). Since only the current in the Ohmic regime
contributes to $S_{env}$, there is an additional high-energy cutoff $\sim {c}^{2}/(4\pi \sigma ^{2D}{d})$.
Therefore, in Eq.~(\ref{eq:tunneling_frequency}) $\tau _{tun}^{-1}$ should be replaced by
$\tilde{\tau}_{tun}^{-1}=\min \{\tau _{tun}^{-1},c^{2}/(4\pi \sigma ^{2D}{d})\}$. Little can be said about
$\tau_{tun}$ as the effective mass of the vortex and the potential of the tunneling barrier depend on the
specific properties of the superconducting film. As to the energy scale $c^{2}/(4\pi \sigma ^{2D}{d})$, it is
evaluated to be $\sim 10^{3}K$ (the resistivity of the gate is about $10\hspace{1mm}\mu \Omega {cm}$).

The temperature enters in Eq.~(\ref{eq:tunneling_frequency}) as a low-energy cutoff because the excitations with
energy smaller than $T$ do not contribute to the action. The dependence of the tunneling rate on other factors
limiting the time of response of the environment can be found from the RG
analysis.~\cite{Schmid1983,AndersonKondo1970b} Apart from the temperature, such factors include the typical
energy mismatch between the minima of the vortex potential $\varepsilon $ and the tunneling rate itself. The
energy mismatch $\varepsilon$ in this problem is equivalent to a magnetic field in the Kondo problem. Moreover,
low-frequency components of the Eddy current with frequencies smaller than $\Gamma _{G}$ cannot develop when the
tunneling events are too frequent. This is why the tunneling rate determines its own renormalization in a
self-consistent way. To include the influence of these factors, one should substitute $T$ in
Eq.~(\ref{eq:tunneling_frequency}), by $T_{max}=max\{T,\Gamma_G,\varepsilon\}$.

For $K_{E}>1$ the temperature dependence given by Eq.~(\ref{eq:tunneling_frequency}) holds for all temperatures
down to zero. Thus, for $\varepsilon \rightarrow 0$ the vortex becomes localized at $T=0$. In the Kondo
problem~\cite{AndersonKondo1970b} this happens for a ferromagnetic sign of the exchange. The localization occurs
because the strong response of the environment blocks the tunneling. In the opposite case, $K_{E}<1,$ the
tunneling rate remains finite when $T,\varepsilon \rightarrow 0$:
\begin{equation}
\Gamma _{G}\sim \tilde{\tau}_{tun}^{-1}\left(\Gamma_{unG}\tilde{\tau}_{tun}\right) ^{1/(1-K_{E})}.
\label{eq:T_0}
\end{equation}
This quantity has the meaning of the Kondo temperature which determines the physics of switching between two
states when both $T$ and $\varepsilon $ are smaller than $\Gamma _{G}.$ Unlike the Kondo problem, the
dimensionless parameters $K_{E}$ and $(\Gamma _{unG}\tau _{tun})$ which determine the tunneling rate are
completely uncorrelated. We are interested in studying the case when the tunneling is small even in the absence
of the gate, i.e., when $\Gamma _{unG}\ll 1/\tau _{tun}$. Then, the tunneling between two states is equivalent
to the Kondo problem in the limit of extreme anisotropy. In this case the renormalization of $K_{E}$ can be
neglected, and the trajectories of the RG phase-diagram are straight lines.

So far, in the discussion of the renormalization of the tunneling rate we ignored the time dependence of the
log-factor in Eq.~(\ref{eq:ActionEnv}). Generally speaking, this is not valid as there can be a big window of
energies, $\tilde{\tau}_{tun}^{-1}>|\omega _{n}|>\tau _{\lambda }^{-1}$, where $\tau _{\lambda }^{-1}=c^{2}/4\pi
\sigma ^{2D}\lambda $. Then, the RG analysis has to be revised. The modified phase diagram is plotted in
Fig.~\ref{fig:phase_diagram}; see Appendix for details. Notice, that the condition for localization (the
inability to tunnel) becomes harder. It is also worth noting that the Delocalization-Localization phase
transition occurs at a finite energy scale.

\begin{figure}[h]
\begin{flushright}\begin{minipage}{0.5\textwidth} \centerline{
    \includegraphics[width=0.9\textwidth]{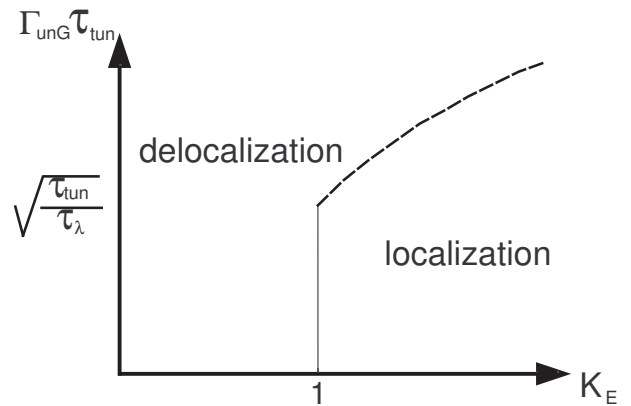}}
\caption{\small The Delocalization-Localization phase diagram for the tunneling of vortices in the plane of
$\Gamma_{unG}\tau_{tun}$ and the dissipation coefficient $K_{E}$.} \label{fig:phase_diagram}
\end{minipage}\end{flushright}
\end{figure}

Let us discuss the result obtained for the exponent $K_{E}$. Naturally, the action $S_{env}$ is proportional to
the square of the magnetic field. After integration over the coordinates one obtains the factor
$(\alpha/\lambda)^{2}$. From the structure of the action~(\ref{eq:ActionEnv}) it follows that $K_{E}$ is
proportional to the squared distance of the change of vortex position, which is typically of order $\xi$.
Altogether, the exponent $K_{E}$ acquires the factor $(\xi /\lambda )^{2}\ll 1$. This smallness is opposed by
the sheet conductance of the gate, $\sigma ^{2D}$, which in the case of a thick well conducting gate gives a
large factor $2\pi \hbar \sigma {d}/e^{2}$. As usually, the effect of dissipation is stronger for cleaner
systems.

\section{Orthogonality Catastrophe~\cite{KM2006}}

\label{sec:OC}

The magnetic flux of a vortex piercing through the gate scatters the electrons in a way similar to Aharonov-Bohm
(\textit{A-B}) scattering.~\cite{Aharonov1959,Aharonov1984} The tunneling of a vortex enforces the electrons
inside the gate to adjust to its new position. The response of the electrons to a sudden change of the vortex
position leads to the \textit{OC} that manifests itself in the vanishing overlap $\langle \Psi _{f}|\Psi
_{i}\rangle $ of the two wave functions describing the macroscopic electron system before and after the change
of the scattering potential.~\cite{AndersonOC1967} The tunneling rate renormalized by the overlap integral is
\begin{equation}
\Gamma _{G}=\Gamma _{unG}\langle \Psi _{f}|\Psi _{i}\rangle. \label{eq:Overlap1}
\end{equation}

The overlap $\langle \Psi _{f}|\Psi _{i}\rangle $ can be expressed in terms of the operators $\hat{S}_{i}$ and
$\hat{S}_{f}$ describing the scattering of the electrons by the magnetic field of the vortex in its initial and
final positions~\cite{Yamada1982}:
\begin{align}\label{eq:Overlap_tunneling}
&|\langle \Psi _{f}|\Psi _{i}\rangle |=N^{-K_{OC}}; \\
&\hspace{1mm}K_{OC}=-\frac{1}{8\pi ^{2}}Tr\left\{ \ln ^{2}(\hat{S}_{f}\hat{S}_{i}^{-1})\right\}.  \notag
\end{align}
Here $N$ is the number of electrons in the gate and, hence, the overlap factor vanishes unless there is a
mechanism that limits the effectiveness of the \textit{OC}. It is clear from the comments to
Eq.~(\ref{eq:tunneling_frequency}) that at finite temperatures~\cite{Yamada1984,Kagan1986} the parameter $1/N$
should be substituted by $(\max\{T,\Gamma _{G}\}\tau _{tun})$. Obviously, the renormalized tunneling rate
$\Gamma_G$ is given by Eqs.~(\ref{eq:tunneling_frequency}) and~(\ref{eq:T_0}) with $K_E$ replaced by $K_{OC}$,
and the localization of vortices can be achieved when $K_{OC}>1$.

In the following part of this section we show that for a superconducting film magnetically coupled to a metallic
gate (see Fig.~\ref{fig:vortex_field}) the exponent $K_{OC}$ is
\begin{equation}
K_{OC}=\varsigma{d}(k_{F}^{gate})^{2}\frac{(\alpha \delta r)^{2}}{64\lambda},
\label{eq:K{OC}}
\end{equation}
where $k_{F}^{gate}$ is the Fermi momentum of the electrons in the gate and the prefactor $\varsigma $ is
evaluated numerically as $\approx 0.4$.

The cylindrical symmetry of the vortex allows us to analyze the scattering of electrons using the basis of
cylindrical waves, $|\ell ,q,k_{z}\rangle $; here $\ell $ is the angular momentum along the $z$ axis, while $q$
and $k_{z}$ are the magnitudes of the in-plane and $z$ components of the momentum. In this basis the elements of
the matrix $S_{f}S_{i}^{-1}$ can be easily calculated in terms of the phase shifts $\delta_{\ell}$ as
\begin{align}\label{eq:Overlap_Matrix_Element}
_{f}\langle {\ell }|S_{f}&S_{i}^{-1}|{\ell ^{\prime }}\rangle _{f} \\\nonumber &=\sum_{n}e^{2i\delta _{\ell
}-2i\delta _{n+\ell }}J_{n}(q\delta r)J_{n-\ell ^{\prime }+\ell }(q\delta r)\ ,
\end{align}
where $J_{\nu }(z)$ is the Bessel function.

To proceed further, we need to find the specific phase shifts for scattering by a vortex. An analogy to
classical scattering, where the angular momentum is related to the impact parameter $b=|\ell |/q,$ helps
elucidate the behavior of the phase shift as a function of $\ell $. For $b\gg {\lambda},$ the scattering by the
vortex is similar to the \textit{A-B} scattering by a flux $\alpha \Phi _{0}$. In the \textit{A-B}
scattering~\cite{Aharonov1959,Aharonov1984} electrons acquire the phase
$\delta_{\ell}^{A-B}=\frac{\pi}{2}(|\ell|-|\ell-\alpha|)$. The uniqueness of this scattering is in its infinite
range: $\delta _{\ell }$ does not vanish when $ |\ell |\rightarrow \infty $. For scattering by the vortex, the
jump in the \textit{A-B} phase shifts is smeared out, but the infinite range character of this scattering is
preserved. Hence, $\delta _{\ell }$ varies monotonically as a function of $\ell $ between the two limits:
\begin{equation}
\delta _{\ell }\xrightarrow[\ell{\gg}q{\lambda}]{}\alpha \frac{\pi }{2}sgn\ \ell.  \label{eq:AB_Phase_Shift}
\end{equation}%
Naturally, for $q{\lambda}\gg 1$ the phase shift depends on $b$ and ${\lambda}$ only through the dimensionless
combination $b/{\lambda}=\ell /q{\lambda}$ such that $\delta _{\ell }=\frac{\alpha \pi }{2}g\left( \ell
/q{\lambda} \right) $; see Fig. $3$ in Ref.~\onlinecite{KM2006} for illustration.

We now notice that the sum determining the elements of $_{f}\langle {\ell }|S_{f}S_{i}^{-1}|{\ell ^{\prime
}}\rangle _{f}$ is accumulated at $-q\delta r\lesssim n\lesssim q\delta r$. This is because the Bessel Functions
$J_{\nu }(z)$ decay exponentially with their order when $\nu >z$. Therefore, since
$\delta{r}/{\lambda}\sim\xi/\lambda\ll1$, the phase shifts difference in Eq.~(\ref{eq:Overlap_Matrix_Element})
can be approximated as:
\begin{equation}
\delta _{\ell }-\delta _{n+\ell }\xrightarrow[n{\ll}q{\lambda}]{}-n\delta
_{\ell }^{\prime };\qquad \delta _{\ell }^{\prime }\approx \frac{\alpha \pi
}{2q{\lambda}}g^{\prime }\left( \frac{\ell }{q{\lambda}}\right) \ll 1.
\label{eq:Phase_shifts_Difference}
\end{equation}%
The final step of the calculation is to expand in $\delta r/{\lambda}$ the logarithm in
Eq.~(\ref{eq:Overlap_tunneling}), and take the trace over $\ell$ and the momentum on the Fermi surface. The
outcome of the calculation is given in Eq.~(\ref{eq:K{OC}}). The gate thickness $d$ appears here as a result of
taking the trace. The specifics of the vortex solution enter only through $g(x)$, with the integral yielding
$\varsigma =\int {dx(dg/dx)}^{2}\approx 0.4$.

Using the known expressions for $\lambda $ and $\xi$ in disordered thin films, the exponent can be rewritten as:
\begin{equation}
K_{OC}\sim \varsigma \frac{\alpha ^{2}}{48\pi }\left( \frac{e^{2}}{{c}}%
\right) ^{2}\frac{v_{F}^{sc}}{e^{2}}%
(k_{F}^{gate}d)(k_{F}^{gate}a)(k_{F}^{sc}l^{sc})^{2}.  \label{eq:K-general}
\end{equation}%
The index $sc$ refers to the electrons in the superconducting film: $l^{sc}$ is their mean free path (in the
normal state) and $v_{F}^{sc}$ is the Fermi velocity. Interestingly, $T_{c}$ drops out from $K_{OC}$ so that it
depends only on the geometrical factors and the non-superconducting properties of electrons. We see that the
value of the exponent $K_{OC}$ is determined by a small factor $\sim 10^{-7}$ opposed by a product of a few
large factors. Unlike $K_{E}$, the condition for vortex localization can be easily fulfilled by $K_{OC}$ for a
not too thin gate and not too disordered superconducting film.

\section{Comparison between the Two Calculations and Discussion}

\label{sec:discussion}

The expressions in Eqs.~(\ref{eq:K{E}}) and (\ref{eq:K{OC}}) for the exponent describing the renormalization of
the tunneling rate have been obtained assuming that only one vortex participates in each tunneling event. In
general, vortices can tunnel as a bundle, or as topological defects. In a vortex lattice or in a glass state
such defects can be dislocations pairs, interstitials or vacancies. Still, the calculation remains valid as long
as $\delta{r}\ll{\lambda}$. This is because the magnetic field of the tunneling vortices extends over a large
distance, so that their exact configurations before and after the tunneling are not important. The only relevant
quantity is the product $\alpha \delta r$. For a single vortex, $\alpha =1/2$. When more than one vortex tunnel
together $\alpha $ should be multiplied by the number of vortices.

The two expressions for the exponent, $K_{E}$ and $K_{OC}$, share the same dependence on $\alpha\delta {r}$ and
$d$. Therefore, the ratio between them is "universal":
\begin{equation}
\frac{K_{E}}{K_{OC}}= \frac{{\sigma}\ln(\lambda/d)}{e^{2}(k_{F}^{gate})^{2}%
\lambda}\sim\frac{l_{gate}}{\lambda},
\end{equation}
where $l_{gate}$ is the mean free path of the electrons in the gate. Since in a thin superconducting film the
penetration depth $\lambda$ is very large, under the conditions of the experiment~\cite{Mason2002}, the
\textit{OC} is dominant.

In order to understand the difference between the two expressions for the exponent, we have to explain the
dependence of $K_{OC}$ on $\lambda$. Although we invoke the expansion in terms of the small parameter $\delta
r/{\lambda}\ll1,$ we get $K_{OC}\propto (\delta r)^{2}/{\lambda}$. This is typical for the \textit{OC} when an
extended scattering potential is considered, because in this case a large number of scattering channels
(harmonics) is involved. Therefore, the total effect of \textit{OC} is parametrically bigger than the one from a
single channel.~\cite{MatveevLarkin1992} For the problem discussed here, the relatively weak dependence of
$K_{OC}$ on $\lambda$ can be understood from the following arguments. It has been shown that the \textit{OC} is
determined by $\sum_{\ell}(\delta_{\ell}-\delta_{\ell+1})^{2}\approx\sum_{\ell}(\delta_{\ell}^{\prime})^{2}$.
Since the phase shifts approach asymptotically the limit $\pm \alpha \frac{\pi }{2}$, the sum
\begin{equation}
\sum_{\ell }(\delta _{\ell }-\delta _{\ell +1})\approx \sum_{\ell }\delta _{\ell }^{\prime }=\pi \alpha.
\label{eq:Sum_of_differnces}
\end{equation}%
Therefore, the result obtained for the exponent $K_{OC}$ corresponds to the differences $(\delta _{\ell }-\delta
_{\ell +1})$ that are distributed almost equally between $L\sim q{\lambda}$ channels:
\begin{equation}
\sum_{\ell }(\delta _{\ell }-\delta _{\ell +1})^{2}\sim L\left( \frac{\pi
\alpha }{L}\right) ^{2}\sim \frac{\alpha ^{2}}{{\lambda}}.
\label{eq:Sum_of_squares}
\end{equation}
Indeed we see that the first power of $\lambda^{-1}$ is natural for the exponent $K_{OC}$ because a large number
of scattering channels is perturbed by the change in the potential when such an extended object as a vortex
tunnels.

The peculiarity of the discussed problem is that a vortex in a thin film is a very extended object. In general,
the tunneling of an extended object excites many channels of the environment. Unlike the standard OC caused by a
scattering potential (but not a vector potential) where the exponent is determined by sum of squares of the
phase shifts, here the sum of the squared derivatives of the phase shifts determines the exponent of the OC.
Still there is some similarity between the two problems. In the standard OC the sum of the phase shifts is
finite because of the Friedel sum rule (see e.g. the tunneling via the localized level considered in
Ref.~\onlinecite{MatveevLarkin1992}), while in the discussed problem the asymptotic limits of the AB phase
shifts make the sum of the derivatives of the phase shifts to be finite, see Eq.~(\ref{eq:Sum_of_differnces}).
As a result of this "AB sum rule" the OC exponent given by Eq.~(\ref{eq:Sum_of_squares}) is similar to that
given by Eq.~(10) in Ref.~\onlinecite{MatveevLarkin1992}.

One may conclude from Eqs.~(\ref{eq:Sum_of_differnces}) and~(\ref{eq:Sum_of_squares}) that the randomization of
the phase shifts due to the disorder can only increase the value of the exponent $K_{OC}$. (In the general case,
$\ell $ should be substituted by the index of the states diagonalizing the scattering matrix.) The scattering by
impurities leads to the randomization of the phase differences, while the asymptotic limits of the phase shifts
remain the same, $\pm \alpha \frac{\pi }{2}$. Therefore, the value of the exponent $K_{OC}$, which is determined
by the squares of the phase differences, should increase in the presence of disorder. This conclusion is in
accordance with the existing theoretical results on the enhancement of the \textit{OC-}exponent by not too
strong disorder.~\cite{Kroha1992, Gefen2002}

The sum over the channels (harmonics) enters in a natural way into the exponent $K_{OC}$, while in the
calculation of the "tunneling with dissipation" the sum is absent. The scheme of calculation of the "tunneling
with dissipation" for the Eddy currents corresponds to the OC expression calculated in the perturbation theory
up to the second order with respect to the change of the vector potential. It is merely finding the elements of
the scattering matrix in the Born approximation. However, the AB effect is a non-perturbative phenomenon. When
the current is calculated in the linear response the asymptotic limits of the phase shifts are not captured. One
may also see that in the linear response scheme only one channel remains in the sum over the channels that is
present in the OC expression. Therefore, the additional factor of $\lambda$ cannot be reproduced within the
"tunneling with dissipation" scheme. This is the reason why $K_{OC}$ is significantly larger than the obtained
$K_{E}$.

The idea to use a double layer system to study the dynamics of vortices is well
known.~\cite{Giaever1965,Klapwijk1991,Rimberg1997} In addition to the magnetic coupling between the film and the
gate, one may consider a capacitive coupling between them. In the case of the Josephson junction arrays (or
granular superconductors) the capacitive coupling reduces the fluctuations of the phase of the superconducting
order parameter.~\cite{Rimberg1997, Wagenblast1997} As a result, the system may undergo a transition from an
insulating to a superconducting state. However, for a homogenous film with a relatively small resistance
$\sim1.5k\Omega /\square$ used in Ref.~\onlinecite{Mason2002} the phase fluctuations are not so
effective.~\cite{Ramakrishnan1989} This is confirmed by the observed insensitivity of the critical magnetic
field $H_{c}$ to the presence of the gate. Furthermore, in homogeneous superconductors the motion of vortices is
not accompanied by the redistribution of the charge density. Therefore, there are good reasons to ignore here
the capacitive coupling between the film and the gate.

\section{Summary}

In this work we have studied how a metallic gate placed above a superconducting film affects the tunneling rate
of the vortices. The gate and the film are coupled by the magnetic field of the vortices that pierces through
the gate. We analyze the renormalization of the tunneling rate by the gate. We consider two approaches to
describe the response of the electrons inside the gate on the tunneling event: (i) the Eddy current in the gate
generated by the motion of vortices, and (ii) the OC caused by the change in the vortex position. The OC is due
to the Aharonov-Bohm scattering of electrons inside the gate. The exponent determining the renormalized
tunneling rate $ \Gamma _{G}(T)=\Gamma _{unG}\left( T\tau _{tun}\right) ^{K}$ is given by Eq.~(\ref{eq:K{E}})
for the effect of the Eddy current, and by Eq.~(\ref{eq:K{OC}}) for the OC. We find that for the experimental
setup of Ref.~\onlinecite {Mason2002} the effect of the OC provides an exponent sufficient for a substantial
suppression of the tunneling rate of the vortices, $K_{OC}\sim 1$.

The peculiarity of the discussed problem results from combination of two elements: the extended size of the
tunneling vortex and the unique features of the AB scattering. In general, the tunneling of an extended object
creates excitations in many channels of the environment. The sum over the channels (harmonics) enters in a
natural way into the exponent describing the effect of the OC, while in the calculation of the "tunneling with
dissipation" the sum is absent. This is the reason why $K_{OC}$  is significantly larger than the obtained
$K_{E}$. The scheme of calculation of the "tunneling with dissipation" for the Eddy currents corresponds to the
OC expression calculated in the perturbation theory up to the second order with respect to the change of the
vector potential. Because of the non-perturbative character of the AB effect, the phase shifts cannot be found
within the Born approximation. Therefore the action $S_{env}$, being formulated in terms of the macroscopic
charge and current densities in the regime of the linear response, is unable to describe the response of all
fluctuation modes that can be excited by a vortex as a result of tunneling.

We address our analysis to a recent experiment~\cite{Mason2002} in which the resistance of a superconducting
film has been measured in a magnetic field both with and without a gate. We interpret the experiment by assuming
that the origin of the resistance at low temperatures is tunneling vortices. From our point of view, the
difference in the resistance of the gated and ungated film indicates that the gate reduces the tunneling rate of
the vortices making them localized. Indeed, we show here that adding a gate may effectively suppress the vortex
tunneling. The gated system discussed here can be used as an effective experimental tool for investigating the
vortex motion at low temperatures. The gated system provides a unique opportunity to study the vortex tunneling
in thin superconducting films by such simple means as varying the characteristics of the gate, in particular the
gate thickness and/or the sheet conductance of the gate. This may help to identify the different mechanisms that
contribute to the suppression of the vortex tunneling rate.

\begin{acknowledgments}
We thank A.~Kapitulnik, T.~M.~Klapwijk, B.~I.~Halperin, D.~E.~Khmelnitskii, A.~D.~Mirlin and B.~Spivak for
useful discussions. AF is supported by the Minerva Foundation.
\end{acknowledgments}

\appendix

\section{Localization-Delocalization Phase Diagram for Eq.~(\ref{eq:ActionEnv})}

The renormalized tunneling rate as given by Eq.~(\ref{eq:tunneling_frequency}) corresponds to the RG equation
\begin{equation}
d\ln (\Gamma _{G}\tau )/d\ln (\tau /\tilde{\tau}_{tun})=(1-K_{E}).
\label{eq:RG stand}
\end{equation}
This equation is valid as long as the logarithmic factor in Eq.~(\ref{eq:ActionEnv}) is independent of the time
difference $|\tau _{1}-\tau _{2}|. $ The time dependence in the logarithmic factor is essential for the energy
interval $\tilde{\tau}_{tun}^{-1}>1/\tau >\tau _{\lambda }^{-1},$ where $\tau _{\lambda }^{-1}=c^{2}/4\pi \sigma
d\lambda.$ Then, the RG equation has to be modified by replacing the dissipation coefficient $K_{E}$ in the
above equation by the energy dependent parameter
$\tilde{K}_{E}=K_{E}[1-\ln(\tau_{\lambda}/\tilde{\tau}_{tun})/\ln(\lambda/d)+\varrho/\ln(\lambda/d)],$ with the
logarithmic variable $\varrho =\ln (\tau /\tilde{\tau}_{tun})$. For the discussed energy interval the "RG
equation" becomes
\begin{equation}
d\ln (\Gamma _{G}\tau )/d\varrho =(1-\tilde{K}_{E}(\varrho ))\label{eq:RG modif}
\end{equation}
One should not be confused with the appearance of $\varrho $ in the RHS of the ''RG equation''. Here we just
integrate $S_{env}$ in the exponent determining the renormalized tunneling rate $\Gamma _{G}\varpropto \exp
(-S_{env}).$ In a sense we calculate something like the Debye-Waller factor created by the dissipative
environment~\cite{Finkel'stein 83}. [The $\varrho $-dependence of the parameter $\tilde{K}_{E}$ can be
reformulated as an RG equation, additional to Eq.~(\ref{eq:RG modif}). Namely,
\begin{subequations}
\begin{equation}
d\tilde{K}_{E}/d\varrho =\left\{\begin{array}{ll} const &
\mbox{\hspace{5mm}$0<\varrho<\ln(\tau_{\lambda}/\tilde{\tau}_{tun})$}
; \\
\hspace{3mm}0 & \mbox{\hspace{5mm}$\varrho>\ln(\tau_{\lambda}/\tilde{
\tau}_{tun})$},%
\end{array}%
\right.  \tag{A2a}
\end{equation}
where $const=K_{E}/\ln (\lambda /d)$. Because of the different dependencies of $\tilde{K}_{E}$ on $\varrho $,
the RG process may develop in two steps. We discuss the details below.]

Localization occurs if in the course of the RG process the tunneling rate $\Gamma _{G}$ decreases faster than
the running scale $\tau ^{-1}$. Hence, the line separating the localized and delocalized states is determined by
the condition $\tilde{K}_{E}(\Gamma_{G})=1$ as long as $\Gamma_{G}>\tau_{\lambda}^{-1}$. Since the expression
for $\tilde{K}_{E}$ assumes that $\varrho <\ln(\tau_{\lambda}/\tilde{\tau}_{tun})$, the modified $\tilde{K}_{E}$
is smaller than $K_{E}$. Therefore, the condition for localization (the inability to tunnel) becomes harder. It
is worth noting that in the discussed problem the phase transition occurs at a finite energy scale
\end{subequations}
\begin{equation}
\ln (1/\Gamma _{G}\tilde{\tau}_{tun})=\frac{\ln (\lambda /d)}{K_{E}}-[\ln
(\lambda /d)-\ln (\tau _{\lambda }/\tilde{\tau}_{tun})].  \tag{A3}
\label{eq: line}
\end{equation}%
To avoid unnecessary complications, from now on we limit ourselves to the case when the energy cutoff $1/\tilde{\tau}_{tun}$ is determined by $%
c^{2}/4\pi \sigma d^{2}$. Then $\ln (\tau _{\lambda }/\tilde{\tau}_{tun})=\ln (\lambda /d)$ and Eq.~(\ref{eq:
line}) reduces to $\ln (1/\Gamma _{G}\tilde{\tau}_{tun})=\ln (\tau _{\lambda }/\tilde{\tau}_{tun})/K_{E}.$ Note
that the line of the phase transition exists for all $K_{E}>1$ and in the delocalized phase $\Gamma _{G}>\tau
_{\lambda }^{-1}.$ Clearly for $K_{E}<1$ the tunneling rate always remains finite.

In order to find the line of the Delocalization-Localization transition in the plane of the dimensionless
parameters $(\Gamma _{unG}\tilde{\tau}_{tun})$ and $K_{E},$ one has to integrate back Eq.~(\ref{eq:RG modif})
starting at $\ln (\Gamma _{G}\tau )=0$. (The value of $\Gamma_{G}$ corresponding to the line of transition
should be found from the condition $\tilde{K}_{E}(\Gamma_{G})=1$). This procedure yields for the boundary
between the two phases
\begin{equation}
\ln (1/\Gamma _{unG}\tilde{\tau}_{tun})=\ln (\tau _{\lambda }/\tilde{\tau}_{tun})/2K_{E}.  \tag{A4}  \label{eq:
line Boundary}
\end{equation}
One may see that along the boundary $(\Gamma _{G}\tilde{\tau}_{tun})=(\Gamma
_{unG}\tilde{\tau}_{tun})^{2}$.

In the delocalized phase, the renormalized tunneling rate can be found by integrating the RG equation starting
from the bare tunneling rate down to the energy when $\ln (\Gamma _{G}\tau )=0$. Since for $K_{E}>1$ the
condition $\ln (\Gamma _{G}\tau )=0$ is satisfied before the running scale $\tau ^{-1}$ reaches $\tau _{\lambda
}^{-1}$, the RG process involves only Eq.~(\ref{eq:RG modif}). For $K_{E}<1$ the situation is more delicate. In
the first step of the RG process Eq.~(\ref{eq:RG modif}) is integrated. If $\tau ^{-1}$ reaches $\tau _{\lambda
}^{-1}$ before $\ln (\Gamma _{G}\tau )=0$ (i.e., $\Gamma _{G}<\tau _{\lambda }^{-1}$), the process should be
continued. (In terms of the bare parameters, this occurs when
$\Gamma_{unG}<\tau_{\lambda}^{-1}(\tau_{\lambda}/\tilde{\tau}_{tun})^{K_{E}/2}$). In the second step one has to
integrate Eq.~(\ref{eq:RG stand}) using $\tau _{\lambda }^{-1}$ as an upper cutoff instead of
$\tilde{\tau}_{tun}^{-1} $ and $\Gamma _{G}(\tau _{\lambda }^{-1})\equiv \Gamma _{\lambda }$ as an initial
value; here $\Gamma_{\lambda}=\Gamma_{unG}(\tilde{\tau}_{tun}/\tau_{\lambda})^{K_{E}/2}$ is the result of the
integration in the previous step for $\Gamma _{G}$. Finally, after the two-steps renormalizion the tunneling
rate is
\begin{equation}
\Gamma _{G}\sim \tau_{\lambda }^{-1}\left( \Gamma _{\lambda }\tau_{\lambda }\right) ^{1/(1-K_{E})}, \tag{A5}
\end{equation}
which is equivalent to
\begin{equation}
\Gamma _{G}\tilde{\tau}_{tun}\sim (\Gamma _{unG}\tau_{tun})^{1/(1-K_{E})}\left( \tilde{\tau}_{\lambda
}/\tilde{\tau}_{tun}\right) ^{K_{E}/2(1-K_{E})}.  \tag{A6}
\end{equation}%
The Delocalization-Localization phase diagram for the tunneling of vortices in the plane of $\Gamma _{unG}\tau
_{tun}$ and the dissipation coefficient $K_{E}$ at zero temperature is presented in
Fig.~\ref{fig:phase_diagram}.

Finally, we touch upon the role of the relativistic effects. In the derivation of the effective
action~(\ref{eq:ActionEnv}) the integration over the momenta has been limited to the Ohmic regime,
$q\gtrsim2\pi\sigma ^{2D}|\omega _{n}|/c^{2}$. We are interested in a gate with high conductivity, such that
$2\pi \sigma ^{2D}/c\gg 1$. In this case there are relativistic effects that have been left out from the action.
As it has been mentioned in Sec.~\ref{sec:EOM}, at high frequencies $|\omega _{n}|/c>q$ the dissipation occurs
through the Cherenkov radiation. The produced dissipation is equivalent to an effective two-dimensional
conductivity that is approximately $c/2\pi.$ Therefore, this source of the dissipation is negligible compared to
the Ohmic dissipation and cannot change our conclusions.


\begin{thebibliography}{99}
\bibitem{Comment} In the case of the quatntum nucleation, the slow stage yields a stronger dependence of the action on the size of the nuclea when it is large.
\end{thebibliography}
\end{document}